\documentclass{appolb}
\usepackage{epsfig,graphicx}
\usepackage{amssymb}

\begin{document}
\title{The Phase Structure of Dense QCD from Chiral Models%
}
\author{Chihiro Sasaki
\address{
Physik Department,
Technische Universit\"{a}t M\"{u}nchen,
D-85747 Garching, Germany
}
}
\maketitle
\begin{abstract}
A new phase of dense QCD proposed in the limit of large number 
of colors, Quarkyonic Phase, is discussed in chiral approaches.
The interplay between chiral symmetry breaking and confinement 
together with the $N_c$ dependence of the phase diagram are 
dealt with in the PNJL model.
We also discuss a possible phase at finite density where chiral 
symmetry is spontaneously broken while its center remains unbroken.
The quark number susceptibility exhibits
a strong enhancement at the restoration point of
the center symmetry rather than that of the chiral symmetry.
This is reminiscent of the quarkyonic transition.
\end{abstract}
  
\section{Introduction}
\label{sec:int}

Model studies of dense baryonic and quark matter have suggested 
a rich phase structure of QCD at temperatures and quark chemical 
potentials being of order $\Lambda_{\rm QCD}$.
Our knowledge on the phase structure is however still limited
and the description of the matter around the phase transitions 
does not reach a consensus because of the non-perturbative nature
of QCD~\cite{qmproc}.

Possible phases and spectra of excitations are guided by
symmetries and their breaking pattern in a medium.
Dynamical chiral symmetry breaking and confinement
are characterized by strict order parameters
associated with global symmetries of the QCD Lagrangian
in two limiting situations:
the quark bilinear $\langle \bar{q}q \rangle$ in the limit
of massless quarks, and the Polyakov loop $\langle \Phi \rangle$ 
in the limit of infinitely heavy quarks.
The system at finite density could also allow other symmetries,
which are not manifest in the QCD Lagrangian but might emerge
in a dense medium.
In this contribution we discuss the phases in dense QCD from
chiral models.

\section{From $N_c=\infty$ to $N_c=3$}
\label{sec:pnjl}

\begin{figure}
\begin{center}
\includegraphics[width=10cm]{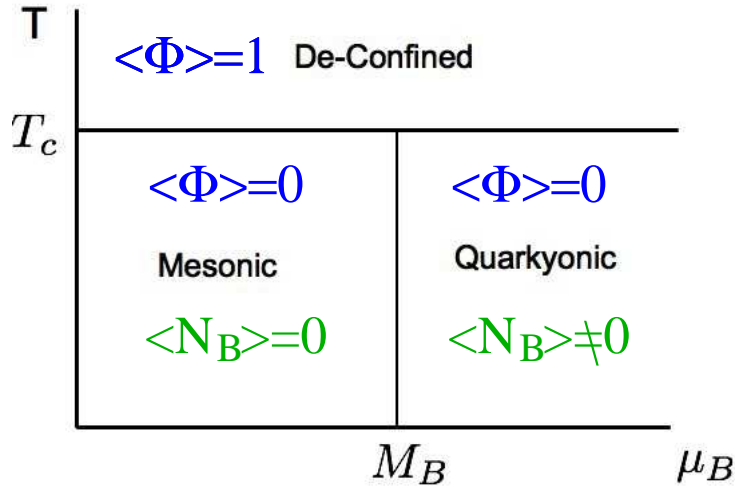}
\caption{
The phase diagram in large $N_c$ proposed in~\cite{quarkyonic}.
}
\label{largeNc}
\end{center}
\end{figure}

A novel phase of dense quarks, Quarkyonic Phase,
was recently proposed based on the argument using 
large $N_c$ counting where $N_c$ denotes number of 
colors~\cite{quarkyonic,larry,rob}:
in the large $N_c$ limit there are three phases which are
rigorously distinguished using $\langle \Phi \rangle$ and the 
baryon number density $\langle N_B \rangle$. The quarkyonic phase 
is characterized by $\langle \Phi \rangle = 0$ indicating the 
system confined and non-vanishing $\langle N_B \rangle$ above 
$\mu_B = M_B$ with a baryon mass $M_B$. The phase structure
in large $N_c$ is shown in Figure~\ref{largeNc}.

\begin{figure}
\begin{center}
\includegraphics[width=10cm]{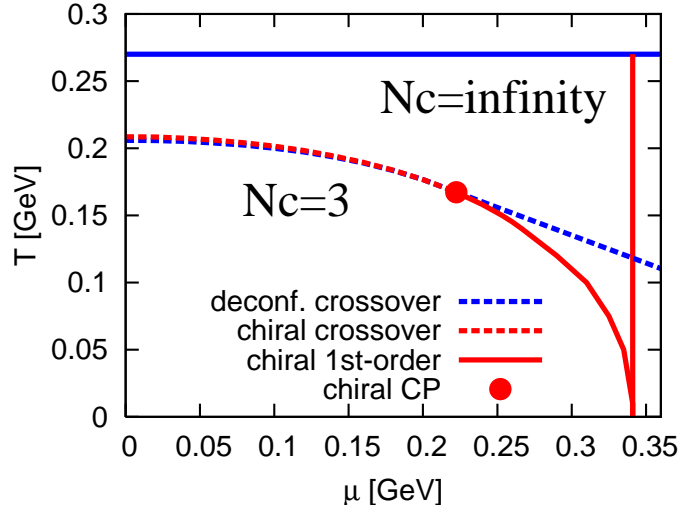}
\caption{
The phase diagram of a PNJL model for different $N_c$~\cite{mrs}. 
Two straight lines indicate the deconfinement and chiral phase 
transitions for $N_c = \infty$ and the lower curves for $N_c=3$.
}
\label{pnjl}
\end{center}
\end{figure}

A possible deformation of the phase boundaries in Figure~\ref{largeNc} 
together with the chiral phase transition can be described
using a chiral model coupled to the Polyakov loop~\cite{mrs}. 
The Nambu--Jona-Lasinio model with Polyakov loops (PNJL model) 
has been developed to deal with chiral dynamics and ``confinement''
simultaneously~\cite{pnjl}.
The model describes that only three-quark states are thermally
relevant below the chiral critical temperature, which is
reminiscent of confinement.
Figure~\ref{pnjl} shows the two transition lines for $N_c=\infty$
and for $N_c=3$ in the two-flavored PNJL model.
In the large $N_c$ limit assuming that the system is
confined, the gap equations for the order parameters 
$\langle \bar{q}q \rangle$ and $\langle \Phi \rangle$ 
become two uncorrelated equations. Consequently,
the quark dynamics carries only a $\mu$ dependence
and the Polyakov loop sector does only a $T$ dependence.
Finite $N_c$ corrections make the transition lines bending down.
The crossover for deconfinement shows a weak dependence on
$\mu$ which is a remnant of the phase structure in large $N_c$.
One finds that for $N_c=3$ deconfinement and chiral crossover lines 
are on top of each other in a wide range of $\mu$. A critical point 
associated with chiral symmetry appears around the junction of those 
crossovers.

The clear separation of the quarkyonic from hadronic phase
is lost in a system with finite $N_c$.
Nevertheless, an abrupt change in the baryon number
density would be interpreted as the quarkyonic transition which 
separates meson dominant from baryon dominant regions.
In fact, a steep increase in the baryon number density and the 
corresponding maximum in its susceptibility $\chi_B$ are driven 
by a phase transition from chirally broken to restored phase 
in most model-approaches using constituent quarks.
One might then consider the chirally symmetric confined phase
as the quarkyonic phase.

The constituent quarks are however unphysical in confined phase.
It is not obvious to have a realistic description of
hadrons from chiral quarks.
In particular, chiral symmetry restoration for baryons must
be worked out. Two alternatives for chirality assignment are 
known~\cite{pdoubling} and it remains an open question which 
scenario is preferred by nature:
(i) in the naive assignment, dynamical chiral symmetry breaking
generates a baryon mass which thus vanishes at the restoration.
(ii) in the mirror assignment, dynamical chiral symmetry breaking
generates a mass difference between parity partners and the chiral
symmetry restoration does not necessarily dictate the chiral
partners being massless. If the chiral invariant mass is not
very small, the baryon number density is supposed to be insensitive
to the quarkyonic transition.

Besides, it seems unlikely that the chirally-restored confined phase
is realized in QCD on the basis of the anomaly matching:
external gauge fields, e.g. photons, interacting with quarks
lead to anomalies in the axial current. Since there are no
Nambu-Goldstone bosons in chiral restored phase, the anomalous
contribution must be generated from the triangle diagram
in which the baryons are circulating. In three flavors, however,
the baryons forming an octet do not contribute to the pole in the 
axial current because of the cancellations~\cite{shifman:anomaly}.
The mirror scenario has nothing to do with this problem because
the sign of the axial couplings to the positive and negative parity 
states are relatively opposite. It is indispensable to any rigorous
argument for this taking account of the physics around the Fermi 
surface, which could lead to a possibility of the chrially restored 
phase with confinement. The anomaly matching conditions at finite 
temperature and density are in fact altered, 
see e.g.~\cite{anomalymatching}.

\section{Role of the tetra-quark at finite density}
\label{sec:znf}

There is a possibility of two different phases with broken
chiral symmetry distinguished by the baryon number density.
An alternative pattern of spontaneous chiral symmetry breaking 
was suggested in the context of QCD at zero temperature and 
density~\cite{stern,SDE,SDE2}.
This pattern keeps the center of chiral group unbroken, i.e.
\begin{equation}
SU(N_f)_L \times SU(N_f)_R \to SU(N_f)_V \times (Z_{N_f})_A\,,
\label{breaking}
\end{equation}
where a discrete symmetry $(Z_{N_f})_A$ is the maximal axial
subgroup of $SU(N_f)_L \times SU(N_f)_R$. The center $Z_{N_f}$ 
symmetry protects a theory from condensate of quark bilinears 
$\langle \bar{q}q \rangle$. Spontaneous symmetry breaking
is driven by quartic condensates which are invariant under
both $SU(N_f)_V$ and $Z_{N_f}$ transformation.
Although meson phenomenology with this breaking pattern
seems to explain the reality reasonably~\cite{stern},
this possibility is strictly ruled out in QCD both at zero
and finite temperatures but at zero density since a different
way of coupling of Nambu-Goldstone bosons to pseudo-scalar
density violates QCD inequalities for density-density
correlators~\cite{shifman}.
However, this does not exclude the unorthodox pattern
in the presence of dense matter. 
In a system with the breaking pattern~(\ref{breaking})
the quartic condensate is the strict order parameter which
separates different chirally-broken phases.

It has been shown that the phase where the symmetry is 
spontaneously broken due to the higher-dimensional operator
is realized as a meta-stable state in an O(2) scalar 
model~\cite{watanabe}.
Another interesting observation came out from
the Skyrme model on crystal:
a new intermediate phase where a skyrmion turns into two half 
skyrmions was numerically found~\cite{skyrmion}.
This phase is characterized by a vanishing quark condensate 
$\langle\bar{q}q \rangle$ and a non-vanishing pion decay constant.
Although the above non-standard pattern of symmetry 
breaking~(\ref{breaking}) was not imposed in the Skyrme Lagrangian,
the result could suggest a dynamical emergence of new symmetries 
in dense environment.

\section{The phase diagram and observables}
\label{sec:phase}

Assuming the symmetry breaking pattern~(\ref{breaking})  
at finite density,
it has been shown that an intermediate phase between
chiral symmetry broken and its restored phases can be realized
using a general Ginzburg-Landau free energy~\cite{hst}.
A 2-quark state $M$ in the fundamental and a 4-quark state $\Sigma$ 
in the adjoint representation are introduced as~\footnote{
 We restrict ourselves to a two-flavor case.
}
\begin{equation}
M_{ij} = \frac{1}{\sqrt{2}}\left( \sigma\delta_{ij} 
{}+ i\phi^a\tau^a_{ij} \right)\,,
\quad
\Sigma_{ab} = \frac{1}{\sqrt{3}}\chi\delta_{ab}
{}+ \frac{1}{\sqrt{2}}\epsilon_{abc}\psi_c\,,
\end{equation}
where the flavor indices run $(i,j) = 1,2$ and
$(a,b,c) = 1,2,3$ and Pauli matrices $\tau^a = 2 T^a$ with
$\mbox{tr}[T^a T^b] = \delta^{ab}/2$.
$\sigma$ and $\chi$ represent scalar fields and 
$\phi$ and $\psi$ pseudoscalar fields, and $\epsilon_{ijk}$
is the total anti-symmetric tensor with $\epsilon_{123} = 1$.
The pion decay constant is read from the Noether current as
\begin{equation}
F_\pi = \sqrt{\sigma_0^2 + \frac{8}{3}\chi_0^2}\,,
\end{equation}
with $\chi_0$ and $\sigma_0$ being the expectation values
of $\chi$ and $\sigma$, determined from the gap equations.
One can deduce a potential in the mean field approximation,
which with an explicit breaking term is obtained as~\footnote{
 A similar potential was considered for a system
 with 2- and 4-quark states under the symmetry breaking
 pattern without unbroken center symmetry in~\cite{rischke} where 
 their 4-quark states are chiral singlet and the potential
 does not include quartic terms in fields.
}
\begin{equation}
V(\sigma,\chi)
=
A\sigma^2 + B\chi^2 + \sigma^4 + \chi^4 - h\sigma
{}+ C\sigma^2\chi + D\chi^3 + F\sigma^2\chi^2\,.
\label{gl}
\end{equation}

\begin{figure}
\begin{center}
\includegraphics[width=10cm]{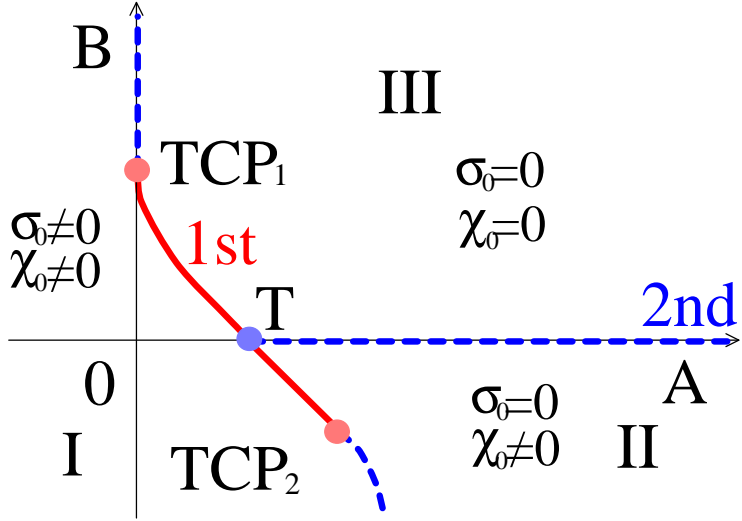}
\caption{
Phase diagram with $C=-1$, $D=F=0$ and $h=0$~\cite{hst}.
The solid and dashed lines indicate first and second order 
phase boundaries, respectively.
One tricritical point, TCP$_1$, is located at
$(A,B)=(0,1/4)$ and another, TCP$_2$, at $(A,B)=(1/4,-1/8)$.
The triple point represented by $T$ is at $(A,B)=(1/8,0)$.
}
\label{D0}
\end{center}
\end{figure}

A phase diagram of this model is shown in Fig.~\ref{D0}.
There are three distinct phases characterized by two order
parameters: Phase I represents the system where both chiral 
symmetry and its center are spontaneously broken due to 
non-vanishing expectation values $\chi_0$ and $\sigma_0$. 
The center symmetry
is restored when $\sigma_0$ becomes zero. However, chiral
symmetry remains broken as long as $\chi_0$ is non-vanishing,
indicated by phase II where the pure 4-quark state is the massless
Nambu-Goldstone boson. The chiral symmetry restoration takes
place under $\chi_0 \to 0$ which corresponds to phase III.
The phases II and III are separated by a second-order line,
while the broken phase I from II or from III is by both first-
and second-order lines. 
Accordingly, 
there exist two tricritical points
(TCPs) and one triple point. One of these TCP, TCP$_2$ 
in Fig.~\ref{D0}, is associated with the center $Z_2$ symmetry
restoration rather than the chiral transition.
The other coefficients $D$ and $F$ change the topology of
the phase diagram and a TCP$_1$ turns to be a critical point
depending on its sign even for $h=0$.

\begin{figure}
\begin{center}
\includegraphics[width=10cm]{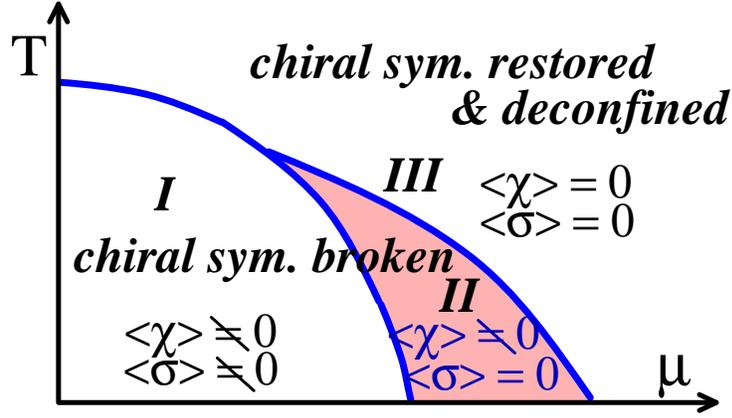}
\caption{
Schematic phase diagram mapped onto $(T,\mu)$ plane.
The lines do not distinguish the order of phase transitions.
The critical point can appear on the boundary that separates
the phase I from III at low $\mu$ and/or the phase I from II 
at intermediate $\mu$~\cite{hst}.
}
\label{phase}
\end{center}
\end{figure}

With an explicit breaking of chiral symmetry
one would draw a phase diagram mapped onto $(T,\mu)$ plane
as in Fig.~\ref{phase}.
The intermediate phase remains characterized
by a small condensation $|\sigma_0| \ll |\chi_0|$.
One would expect a new critical point associated with
the restoration of the center symmetry, CP$_2$, rather
than that of the chiral symmetry if dynamics prefers
a negative coefficient of the cubit term in $\chi$.
Multiple critical points in principle can be observed
as singularities of the quark number susceptibility.

It has been suggested that a similar critical 
point in lower temperature could appear in the QCD phase diagram
based on the two-flavored Nambu--Jona-Lasinio model with vector 
interaction~\cite{KKKN} and a Ginzburg-Landau potential with the 
effect of axial anomaly in three flavors~\cite{yamamoto}.
There the interplay between the chiral (2-quark) condensate and 
BCS pairings plays an important role.
In our framework without diquarks, the critical point is 
driven by the interplay between the 2-quark and
4-quark condensates, where anomalies have nothing to do with 
its appearance.
Besides, the universality class which the critical point in our model 
belongs to is expected to be different from the anomaly-induced
one since spontaneous breaking of $U(1)_B$ is not imposed 
in~(\ref{breaking}).

\begin{figure}
\begin{center}
\includegraphics[width=10cm]{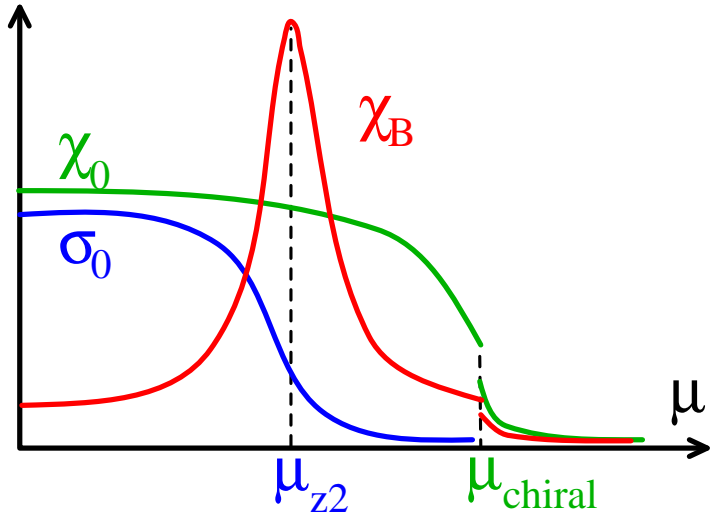}
\caption{
The behavior of the baryon number susceptibility
as a function of chemical potential.
}
\label{qsus}
\end{center}
\end{figure}

Appearance of the above intermediate phase seems to have
a similarity to the notion of Quarkyonic Phase. 
The transition from hadronic to quarkyonic world
can be characterized by a rapid change in the net baryon number
density. In our model this feature is driven by the restoration 
of center symmetry and is due to the fact that the Yukawa coupling 
of $\chi$ to baryons is not allowed by the $Z_2$ invariance.
Fig.~\ref{qsus} shows an expected behavior of
the quark (baryon) number susceptibility which exhibits
a maximum when across the $Z_2$ cross over.
This can be interpreted as the realization  
of the quarkyonic transition in $N_c=3$ world.
How far $\mu_{z2}$ from $\mu_{\rm chiral}$ is depends
crucially on its dynamical-model description.
Thus, the present analysis does not exclude the possibility
that both transitions take place simultaneously and in such
case enhancement of $\chi_B$ is driven by chiral phase transition.
The phase with $\chi_0\neq0$ and $\sigma_0=0$ does not seem to
appear in the large $N_c$ limit~\cite{SDE2,shifman,watanabe}.
It would be expected that including $1/N_c$ corrections induce 
a phase with unbroken center symmetry.

\section{Conclusions}
\label{sec:conclusions}

We have discussed the phases in dense QCD from chiral approaches
along with the anomaly matching which is a field-theoretical
requirement. A possibility of a non-standard breaking pattern leads 
to a new phase where chiral symmetry is spontaneously broken 
while its center symmetry is restored. This might appear as an 
intermediate phase between chirally broken and restored phases 
in $(T,\mu)$ plane. The appearance of this phase also suggests 
a new critical point in low temperatures.
A tendency of the center symmetry restoration is carried by
the net baryon number density which shows a rapid increase
indicating baryons more activated,
and this is reminiscent of the quarkyonic transition.
Dynamical breaking of chiral symmetry $SU(N_f)_L \times SU(N_f)_R$ 
down to $SU(N_f)_V \times {(Z_{N_f})}_A$ should be addressed
in microscopic calculations using the Swinger-Dyson equations or
Nambu--Jona-Lasinio type models with careful treatment
of the quartic operators. 
The properties of baryons near the chiral phase transition
are also an issue to be clarified. Depending on the chirality
assignment to baryons, equations of state may be altered.
In this respect, it attracts an interest
that a top-down holographic QCD model predicts the same sign of 
the axial couplings to the parity partners~\cite{hss}, 
i.e. the naive scenario seems to be preferred.

\section*{Acknowledgments}

I am grateful for fruitful collaboration with 
M.~Harada, L.~McLerran, K.~Redlich and S.~Takemoto.
The work has been supported in part by the DFG cluster 
of excellence ``Origin and Structure of the Universe''.


\end{document}